\documentclass[nofootinbib,aps,11pt]{revtex4-1}
\usepackage{graphicx}
\usepackage{hyperref}
\usepackage{cancel}
\usepackage{amssymb}
\usepackage{textcomp}
\usepackage{amsmath}
\usepackage{bm}
\usepackage{times}
\usepackage{epsfig}
\usepackage{color}

\begin{document}
\title{\Large Single Production of Doubly Charged Higgs at Muon Collider}
\bigskip
\author{Jie-Cheng Jia$^1$}
\author{Zhi-Long Han$^1$}
\email{sps\_hanzl@ujn.edu.cn}
\author{Fei Huang$^1$}
\email{sps\_huangf@ujn.edu.cn}
\author{Yi Jin$^{1,2}$}
\author{Honglei Li$^1$}
\email{sps\_lihl@ujn.edu.cn}
\affiliation{
$^1$School of Physics and Technology, University of Jinan, Jinan, Shandong 250022, China
\\
$^2$Guangxi Key Laboratory of Nuclear Physics and Nuclear Technology, Guangxi Normal University, Guilin, Guangxi 541004, China
}
\date{\today}

\begin{abstract}
	In this paper, we study the single production of doubly charged Higgs $H^{\pm\pm}$ in the type-II seesaw at the high energy muon collider. Compared with the pair production channel $\mu^+\mu^-\to H^{++} H^{--}$, the single production channel $\mu^+\mu^-\to \mu^\mp\ell^\mp H^{\pm\pm}$ in principle could probe the mass region above the threshold $m_{H^{\pm\pm}}>\sqrt{s}/2$. The single production channel depends on the Yukawa coupling $h$, which is related to the neutrino oscillation parameters. We show that the Majorana phases $\phi_1$ and $\phi_2$ have great impact on the individual cross section of the single production. We find that the same sign dilepton signature from $H^{\pm\pm}\to \ell^\pm\ell^\pm$ could probe $m_{H^{\pm\pm}}\lesssim2.6(7.1)$ TeV at the 3 (10) TeV muon collider when the triplet VEV $v_\Delta\lesssim3$ eV.

\end{abstract}

\maketitle

\section{Introduction}

The tiny neutrino masses are implied by the neutrino oscillation experiments \cite{Super-Kamiokande:1998kpq,SNO:2002tuh} together with the cosmological observation \cite{Planck:2018vyg}. Although various pathways are proposed to explain the origin of neutrino masses \cite{Boucenna:2014zba}, the ultimate mechanism needs further experimental conformation \cite{Deppisch:2015qwa}. The extensively studied type-I seesaw mechanism introduces heavy Majorana neutrinos \cite{Minkowski:1977sc,Mohapatra:1979ia,Schechter:1980gr}, whose masses are far beyond the scope of current experiments if large Yukawa couplings are required. An attractive low-scale mechanism is the type-II seesaw \cite{Magg:1980ut,Cheng:1980qt,Lazarides:1980nt,Mohapatra:1980yp}, where the tiny neutrino masses originate from the small vacuum  expectation value (VEV) of triplet scalar $v_\Delta$.

Phenomenological studies of type-II seesaw have been carried out in various literature \cite{FileviezPerez:2008jbu,Akeroyd:2009hb,Akeroyd:2010ip,Akeroyd:2011zza,Aoki:2011pz,Chun:2012zu,Han:2015hba,Han:2015sca,Mitra:2016wpr,Li:2018jns,Dev:2018sel,Du:2018eaw,Padhan:2019jlc,Chun:2019hce,Fuks:2019clu,Yang:2021skb,Ashanujjaman:2023tlj,Banerjee:2024jwn,Bolton:2024thn}. One smoking gun signature is the production of doubly charged Higgs $H^{\pm\pm}$ at colliders \cite{Melfo:2011nx,Cai:2017mow}, then the leptonic decay mode $H^{\pm\pm}\to \ell^\pm\ell^\pm$ leads to the clear signal of lepton number violation. Moreover, the structure of the Yukawa coupling $h$ involved in this decay is determined by the neutrino oscillation parameters \cite{Akeroyd:2007zv,Mandal:2022ysp,Mandal:2022zmy}. By measuring the branching ratio of $H^{\pm\pm}\to \ell^\pm\ell^\pm$, we can even probe the additional Majorana phases, which are insensitive at oscillation experiments. 

After the proposal of the muon collider, searches of doubly charged Higgs at muon collider have  been investigated \cite{Bai:2021ony,Li:2023ksw,Maharathy:2023dtp,Jueid:2023qcf,Belfkir:2023lot}.  Meanwhile, the same-sign $\mu^+\mu^+$ collider is also viable to probe the doubly charged Higgs \cite{Yang:2023ojm,Fridell:2023gjx,Lichtenstein:2023iut,Dev:2023nha}, where the lepton flavor violation processes as $\mu^+\mu^+\to\mu^+\tau^+$ mediated by $H^{++}$ have particular interest. For the pair production of doubly charged Higgs at muon collider, the promising region is restricted by the threshold $ m_{H^{\pm\pm}}<\sqrt{s}/2$. With relatively large Yukawa couplings, another pathway to probe the double charged Higgs is via the single production channel at lepton colliders \cite{Godfrey:2001xb,Yue:2010zu,Montalvo:2012qg,Das:2023tna}, which is hopeful to test the mass region above the pair production threshold $m_{H^{\pm\pm}}>\sqrt{s}/2$ \cite{Xu:2023ene}. 

In this paper, we study the single production of doubly charged Higgs at the TeV scale muon collider, i.e., $\mu^+\mu^-\to \mu^\mp\ell^\mp H^{\pm\pm}$. Previous studies usually fix the oscillation parameters to the best-fit values and vanishing Majorana phases for illustration \cite{Yue:2010zu,Dev:2023nha}.  The production cross section of this process depends on the Yukawa coupling $h$, thus the neutrino oscillation parameters and triplet VEV $v_\Delta$ could impact. The dependence of the cross section on the oscillation parameters is further studied in this paper.  The leptonic decay $H^{\pm\pm}\to \ell^\pm\ell^\pm$ then leads to the same sign dilepton signature $\ell^\pm\ell^\pm$. We also investigate the sensitivity of this signature at the TeV scale muon collider.

The rest of this paper is organized as follows. In Section \ref{SC:MD}, we review the type-II seesaw model and discuss relevant experimental constraints. In Section \ref{SC:DC}, we review the leptonic decay of doubly charged Higgs. Single production of doubly charged Higgs at muon collider is considered in Section \ref{SC:PD}. The sensitivity of the same sign  dilepton signature is searched in Section \ref{SC:SG}. Conclusion is in Section \ref{SC:CL}.

\section{Type-II Seesaw}\label{SC:MD}

In addition to the standard Higgs doublet $\Phi$, a scalar triplet with hypercharge $Y=2$ is introduced in the type-II seesaw as
\begin{equation}
	\Delta=\left(\begin{array}{cc}
		\Delta^{+}/\sqrt{2} & \Delta^{++} \\
		\Delta^{0} & -\Delta^{+}/\sqrt{2}
	\end{array}\right).
\end{equation}

The scalar potential of the type-II seesaw is 
\begin{equation}
	\begin{aligned}\label{Eq:VPhi}
		V(\Phi, \Delta) &=-m_{\Phi}^{2} \Phi^{\dagger} \Phi+M^{2}_\Delta \operatorname{Tr}\left(\Delta^{\dagger} \Delta\right)+\left(\mu \Phi^{\mathrm{T}} i\tau_{2} \Delta^{\dagger} \Phi+\mathrm{h.c.}\right)+\frac{\lambda_0}{4}\left(\Phi^{\dagger} \Phi\right)^{2} \\
		&+\lambda_{1}\left(\Phi^{\dagger} \Phi\right) \operatorname{Tr}\left(\Delta^{\dagger} \Delta\right)+\lambda_{2}\left[\operatorname{Tr}\left(\Delta^{\dagger} \Delta\right)\right]^{2}+\lambda_{3} \operatorname{Tr}\left[\left(\Delta^{\dagger} \Delta\right)^{2}\right]+\lambda_{4} \Phi^{\dagger} \Delta \Delta^{\dagger} \Phi,
	\end{aligned}
\end{equation}
where $\tau_{2}$ is the second Pauli matrix. After the spontaneous symmetry breaking, seven physical scalars are left, i.e., doubly charged Higgs $H^{\pm\pm}$, singly charged Higgs $H^\pm$, CP-even Higgs bosons $h^0$ and $H^0$, and CP-odd Higgs $A^0$ \cite{Arhrib:2011uy}. For simplicity, a degenerate mass spectrum $m_{H^{\pm\pm}}=m_{H^\pm}=m_{H^0}=m_{A^0}$ is assumed in this paper. The collider limits depend on the VEV of the triplet scalar \cite{Ashanujjaman:2021txz}. When $v_\Delta<10^{-4}$~GeV, the ATLAS experiment could exclude $m_{H^{\pm\pm}}<1080$ GeV in the dilepton channel $H^{\pm\pm}\to \ell^\pm\ell^\pm$ \cite{ATLAS:2022pbd}. On the other hand, the upper limit is about 350 GeV in the diboson channel $H^{\pm\pm}\to W^\pm W^\pm$ when $v_\Delta > 10^{-4}$~GeV\cite{ATLAS:2021jol}.

The lepton number violation $\mu$-term in Eqn. \eqref{Eq:VPhi} induces a small VEV of triplet scalar as
\begin{equation}
	v_\Delta \approx \frac{\mu v_\Phi^2}{\sqrt{2} M^2_\Delta}.
\end{equation}

The Yukawa interaction that generates tiny neutrino mass is given by
\begin{equation}
	\mathcal{L}_{Y}= h \overline{L_{L}^{c}} i \tau_{2} \Delta L_{L}+\text { h.c.},
	\end{equation}
where $L_L$ is the left-handed lepton doublet. The neutrinos then obtain masses as
\begin{equation}
	m_\nu = \sqrt{2} h v_\Delta.
\end{equation} 
In this paper, we consider eV scale $v_\Delta$ with relatively large Yukawa coupling $h$. Using neutrino oscillation parameters, the coupling $h$ can be expressed as
\begin{equation}\label{Eq:h}
	h=\frac{U \hat{m}_\nu U^T}{\sqrt{2} v_\Delta},
\end{equation}
where $\hat{m}_\nu=\text{diag}(m_1,m_2,m_3)$ is the diagonalized neutrino mass. The mixing matrix $U$ is denoted as
\begin{equation}
	U = \begin{pmatrix}
		c_{12}c_{13} & s_{12}c_{13} & s_{13}e^{-i\delta_{\text{CP}}} \\
		-s_{12}c_{23} - c_{12}s_{23}s_{13}e^{i\delta_{\text{CP}}} & c_{12}c_{23} - s_{12}s_{23}s_{13}e^{i\delta_{\text{CP}}} & s_{23}c_{13} \\
		s_{12}s_{23} - c_{12}c_{23}s_{13}e^{i\delta_{\text{CP}}} & -c_{12}s_{23} - s_{12}c_{23}s_{13}e^{i\delta_{\text{CP}}} & c_{23}c_{13}
	\end{pmatrix} \times  \text{diag}(1, e^{i\phi_1/2}, e^{i\phi_2/2} ), 
\end{equation}
where $s_{ij}\equiv \sin\theta_{ij}$, $c_{ij}\equiv \cos\theta_{ij}$, $\delta_{\text{CP}}$ is the Dirac phase, and $\phi_1,\phi_2$ are the two Majorana phases.

The scalar triplet could induce lepton flavor violation processes. The branching ratio of $\mu\to e\gamma$ is calculated as \cite{Akeroyd:2009nu}
\begin{equation}
	\text{BR}(\mu\to e\gamma) = \frac{27 \alpha |(h^\dag h)_{e\mu}|^2}{64\pi G_F^2 m_{H^{++}}^4},
\end{equation}
where $\alpha=1/137$ is the fine structure constant, $G_F$ is the Fermi constant. The current MEG limit is BR$(\mu\to e\gamma)<4.2\times10^{-13}$ \cite{MEG:2016leq}, which typically requires $m_{H^{++}} v_\Delta \gtrsim0.78~\text{TeV}\cdot \text{eV}$.

\section{Decay Property of $H^{\pm\pm}$}\label{SC:DC}

Decays of doubly charged Higgs depend on the value of $v_\Delta$ and the mass spectrum of triplet scalars \cite{Melfo:2011nx}. To obtain observable single production of doubly charged Higgs at muon collider, an eV scale $v_\Delta$ is required. For $v_\Delta<10^{-4}$ GeV with the degenerate mass spectrum, the dominant decay mode is $H^{++}\to \ell^+\ell^+$. The partial decay width of the same sign dilepton channel is calculated as
\begin{equation}
	\Gamma(H^{++}\to\ell_i^+\ell_j^+)=\frac{m_{H^{++}}}{4\pi(1+\delta_{ij})} |h_{ij}|^2,
\end{equation} 
where $h_{ij}$ is determined by the neutrino oscillation parameters with Eqn~\eqref{Eq:h}. The explicit expressions for $h_{ij}$ are
\begin{eqnarray}\label{Eq:hij}
	h_{ee} & = & \frac{1}{\sqrt{2} v_\Delta}\Big(m_1 c_{12}^2 c_{13}^2+m_2 s_{12}^2 c_{13}^2 e^{i\phi_1}+m_3 s_{13}^2 e^{-2i\delta_{\text{CP}}} e^{i\phi_2}\Big), \\ \nonumber
	h_{e\mu} & = & \frac{1}{\sqrt{2} v_\Delta} \Big(m_1 (-s_{12} c_{23}-c_{12} s_{23} s_{13} e^{i\delta_{\text{CP}}}) c_{12} c_{13} \\ \nonumber
	& & + m_2 (c_{12} c_{23} - s_{12} s_{23} s_{13} e^{i \delta_{\text{CP}}}) s_{12} c_{13} e^{i \phi_1}+ m_3 s_{23} c_{13} s_{13} e^{-i \delta_{\text{CP}}} e^{i\phi_2} \Big), \\ \nonumber
	h_{e\tau} & = & \frac{1}{\sqrt{2} v_\Delta} \Big( m_1 (s_{12} s_{23} - c_{12} c_{23} s_{13} e^{i\delta_{\text{CP}}}) c_{12} c_{13} \\ \nonumber
	& & + m_2 (-c_{12} s_{23}-s_{12}c_{23}s_{13}e^{i\delta_{\text{CP}}})s_{12}c_{13}e^{i\phi_1}
	+ m_3 c_{23}c_{13}s_{13}e^{-i\delta_{\text{CP}}}e^{i\phi_2} \Big), \\ \nonumber
	h_{\mu \mu} & = & \frac{1}{\sqrt{2} v_\Delta} \Big(m_1(\!-\!s_{12}c_{23}\!-\!c_{12}s_{23}s_{13}e^{i\delta_{\text{CP}}})^2  \!+\!m_2(c_{12}c_{23}\!-\!s_{12}s_{23}s_{13}e^{i\delta_{\text{CP}}})^2 e^{i\phi_1}\!+\! m_3 s_{23}^2 c_{13}^2 e^{i\phi_2} \Big), \\ \nonumber
	h_{\mu \tau} & = & \frac{1}{\sqrt{2} v_\Delta} \Big(m_1(-s_{12}c_{23}-c_{12}s_{23}s_{13}e^{i\delta_{\text{CP}}}) (s_{12} s_{23} - c_{12} c_{23} s_{13} e^{i\delta_{\text{CP}}})\\ \nonumber
	& & + m_2 (c_{12}c_{23}\!-\! s_{12}s_{23}s_{13}e^{i\delta_{\text{CP}}}) (\!-\!c_{12} s_{23}\!-\!s_{12}c_{23}s_{13}e^{i\delta_{\text{CP}}}) e^{i\phi_1}\!+\!m_3 c_{23}s_{23}c_{13}^2 e^{i\phi_2}\Big), \\ \nonumber
	h_{\tau\tau} & =& \frac{1}{\sqrt{2} v_\Delta} \Big( m_1 (s_{12} s_{23}\! -\! c_{12} c_{23} s_{13} e^{i\delta_{\text{CP}}})^2  \!+\! m_2 (\!-\!c_{12} s_{23}\!-\!s_{12}c_{23}s_{13}e^{i\delta_{\text{CP}}})^2 e^{i\phi_1}
	\!+\! m_3 c_{23}^2 c_{13}^2 e^{i\phi_2}\Big).
\end{eqnarray}

Considering the fact that at the time of muon collider operation, the neutrino oscillation parameters might be well measured. In the following study, we fix the measured oscillation parameters to their best fit values at present \cite{Esteban:2020cvm}. Then the free parameters are the lightest neutrino mass $m_{1}(m_3)$, Dirac phase $\delta_{\text{CP}}$, and  two Majorana phases $\phi_1,\phi_2$.

\begin{figure}
	\begin{center}
		\includegraphics[width=0.45\linewidth]{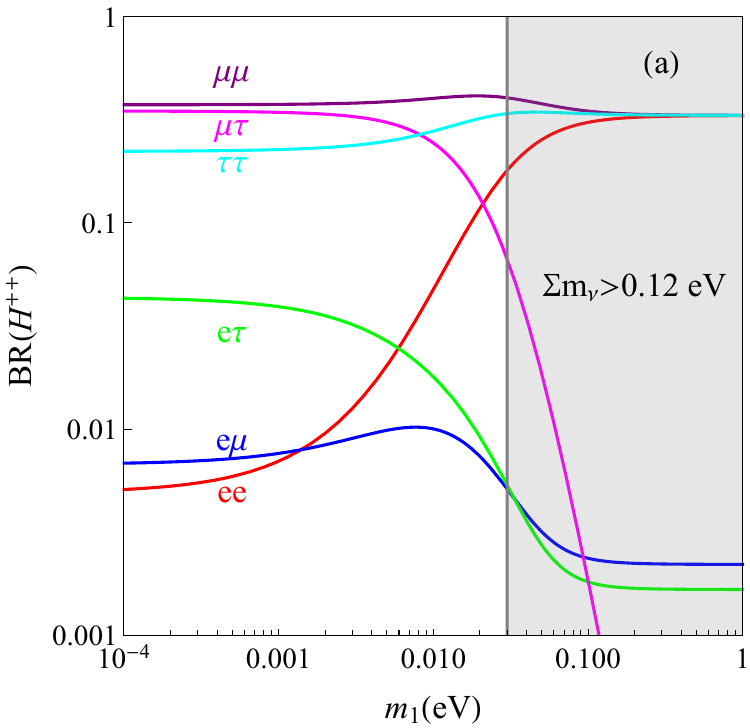}
		\includegraphics[width=0.45\linewidth]{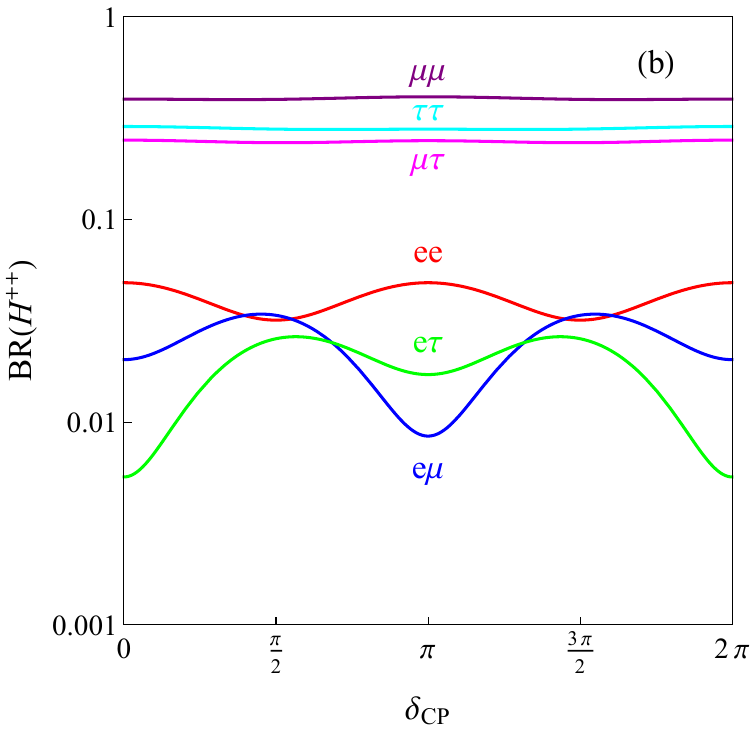}
		\includegraphics[width=0.45\linewidth]{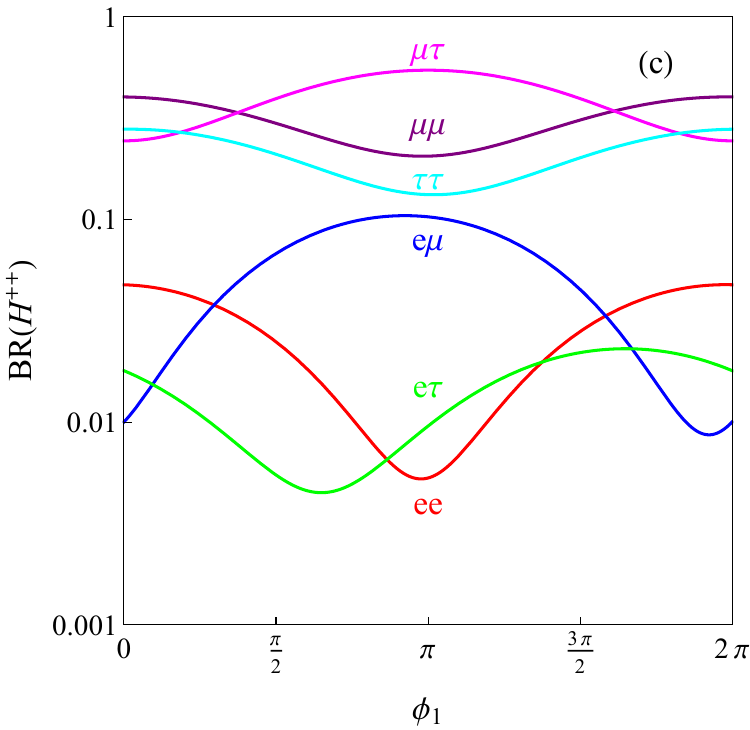}
		\includegraphics[width=0.45\linewidth]{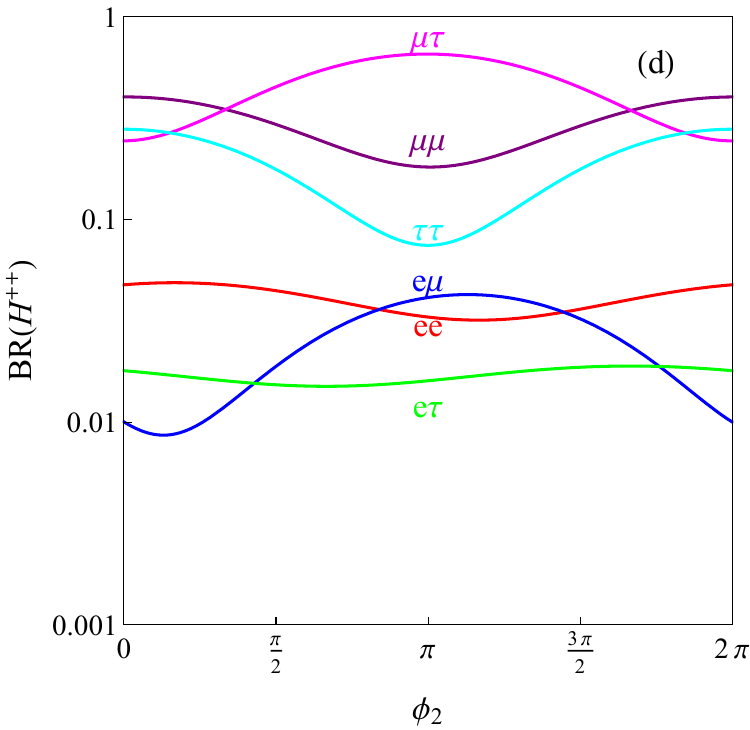}
	\end{center}
	\caption{Branching ratio of $H^{++}$ in NH. In panel (a), we fix $\delta_{\text{CP}}=197^{\circ}, \phi_1=\phi_2=0$. The gray region is excluded by the Planck limit $\sum m_\nu <0.12$ eV \cite{Planck:2018vyg}. In panel (b), we fix $m_1=0.01$ eV, $\phi_1=\phi_2=0$. In panel (c), we fix $m_1=0.01$ eV, $\delta_{\text{CP}}=197^{\circ}$, $\phi_2=0$. In panel (d), we fix $m_1=0.01$ eV, $\delta_{\text{CP}}=197^{\circ}$, $\phi_1=0$. }
	\label{BRNH}
\end{figure}

We first illustrate the branching ratio of $H^{++}$ in the  normal hierarchy (NH), which is shown in Figure \ref{BRNH}. The $H^{++}\to \mu^+\mu^+,\tau^+\tau^+$ channels are the dominant decay modes. For the lightest neutrino mass $m_1<10^{-3}$ eV, the impact of $m_1$ would be small. On the other hand, $m_1$ can be determined by precise measurement of BR$(H^{++}\to \mu^+\tau^+,e^+\tau^+,e^+e^+)$, if $m_1>10^{-3}$ eV. It is clear in panel (b) of Figure \ref{BRNH} that the dominant channels $H^{++}\to \mu^+\mu^+,\tau^+\tau^+,\mu^+\tau^+$ are not sensitive to the Dirac phase $\delta_{\text{CP}}$. Using the subdominant channels $H^{++}\to e^+e^+,e^+\mu^+,e^+\tau^+$, it might be possible to probe $\delta_{\text{CP}}$. The Majorana phases $\phi_1$ and $\phi_2$ have a significant impact on all decay modes of $H^{++}$. For instance, the lepton flavor violation mode $H^{++}\to \mu^+\tau^+$ could be larger than the lepton flavor conserving modes $H^{++}\to \mu^+\mu^+,\tau^+\tau^+$ when $\phi_1=\pi, \phi_2=0$ or $\phi_1=0,\phi_2=\pi$.

\begin{figure}
	\begin{center}
		\includegraphics[width=0.45\linewidth]{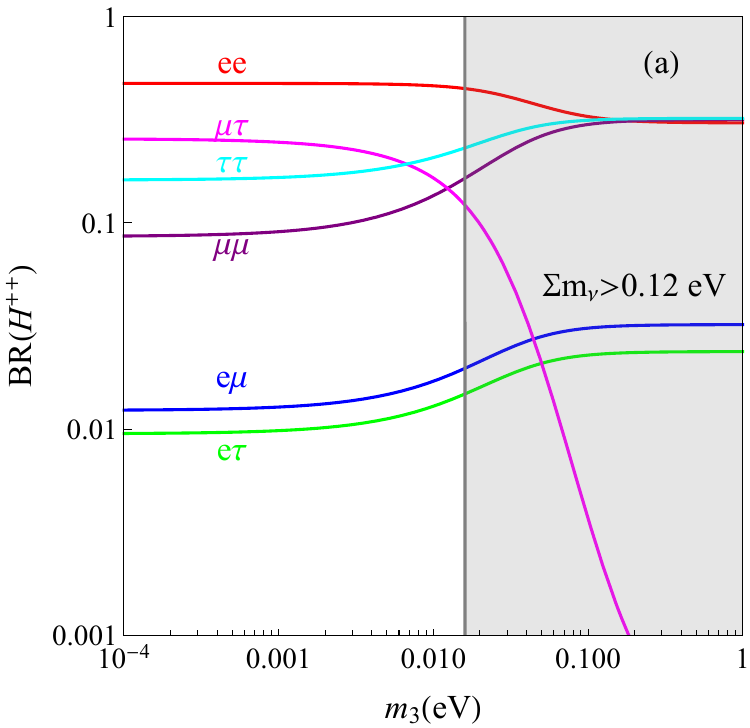}
		\includegraphics[width=0.45\linewidth]{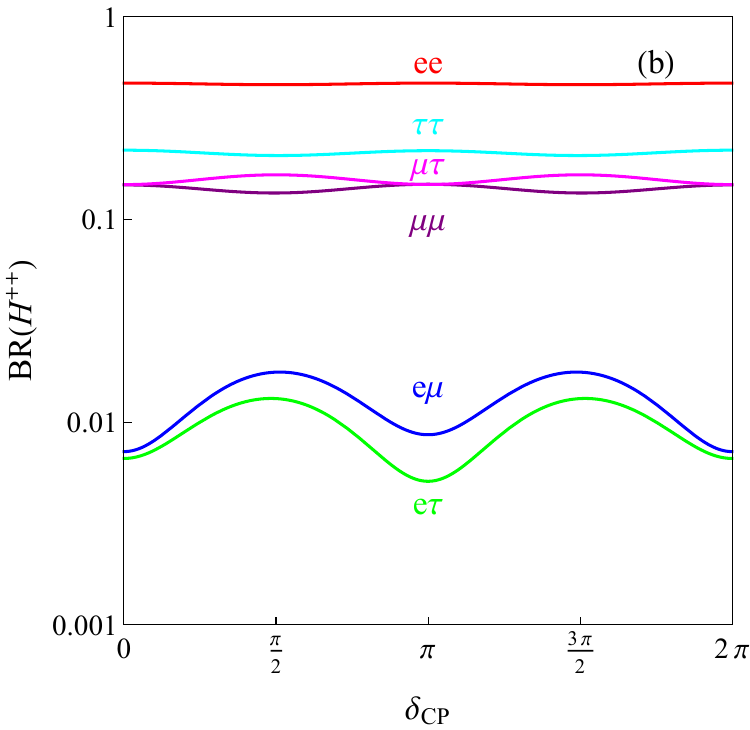}
		\includegraphics[width=0.45\linewidth]{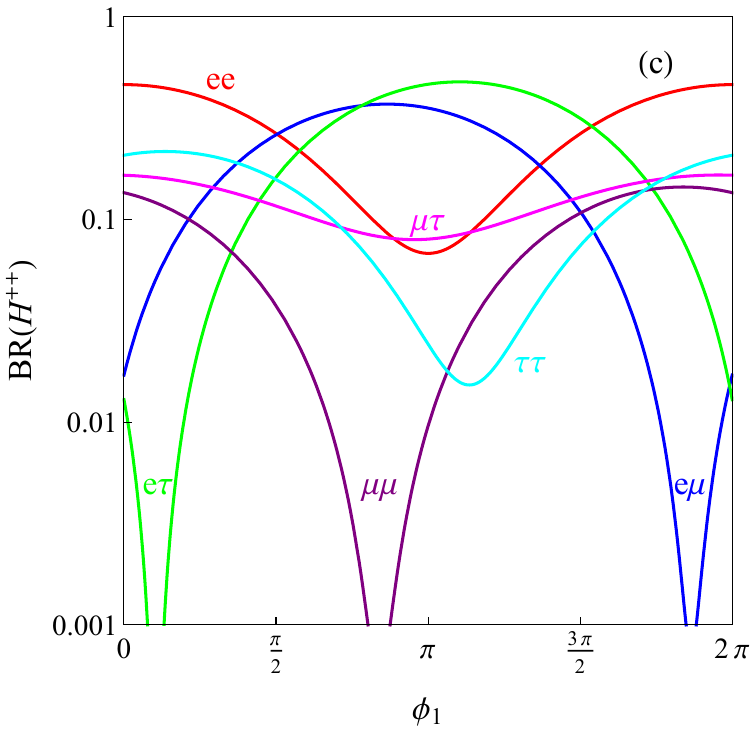}
		\includegraphics[width=0.45\linewidth]{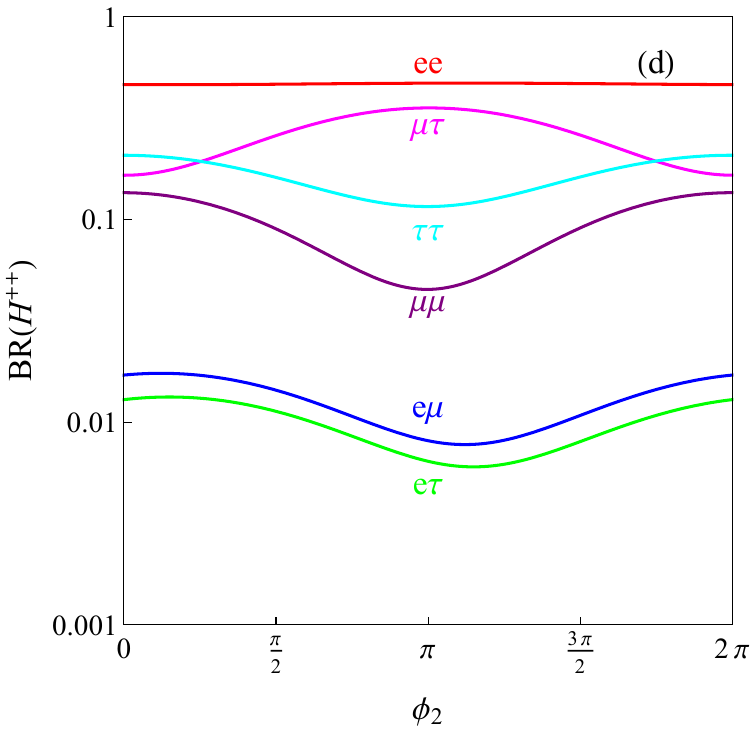}
	\end{center}
	\caption{Branching ratio of $H^{++}$ in IH. In panel (a), we fix $\delta_{\text{CP}}=282^{\circ}, \phi_1=\phi_2=0$. The gray region is excluded by the Planck limit $\sum m_\nu <0.12$ eV \cite{Planck:2018vyg}. In panel (b), we fix $m_1=0.01$ eV, $\phi_1=\phi_2=0$. In panel (c), we fix $m_1=0.01$ eV, $\delta_{\text{CP}}=282^{\circ}$, $\phi_2=0$. In panel (d), we fix $m_1=0.01$ eV, $\delta_{\text{CP}}=282^{\circ}$, $\phi_1=0$.  }
	\label{BRIH}
\end{figure}

The results of BR$(H^{++})$ in the inverted hierarchy (IH) scenario are shown in Figure ~\ref{BRIH}. In this case, the $H^{++}\to e^+e^+$ channel becomes the dominant one. Under the constraint from cosmology of $\sum m_\nu<0.12$~eV, we might probe the lightest neutrino mass $m_3$ by precise measurement of BR($H^{++}\to \mu^+\tau^+$). The $H^{++}\to e^+e^+$ mode is not sensitive to the Dirac phase $\delta_{\text{CP}}$ and Majorana phase $\phi_2$. Sizable deviations happen in the $H^{++}\to \mu^+\mu^+,\mu^+\tau^+$ modes when varying $\delta_{\text{CP}}$. The Majorana phase $\phi_1$ has a larger impact than the Majorana phase $\phi_2$. For example, varying $\phi_1$ could lead to BR$(H^{++}\to \mu^+\mu^+)<10^{-3}$, while  BR$(H^{++}\to \mu^+\mu^+)>0.04$ is always satisfied when varying $\phi_2$.

\begin{figure}
	\begin{center}
		\includegraphics[width=0.45\linewidth]{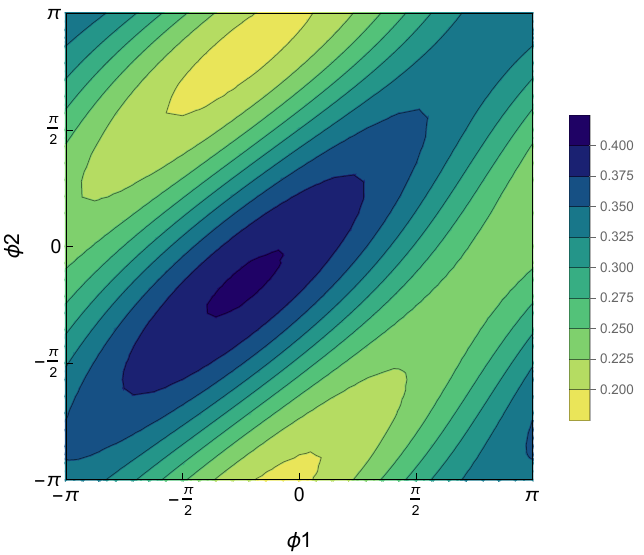}
		\includegraphics[width=0.45\linewidth]{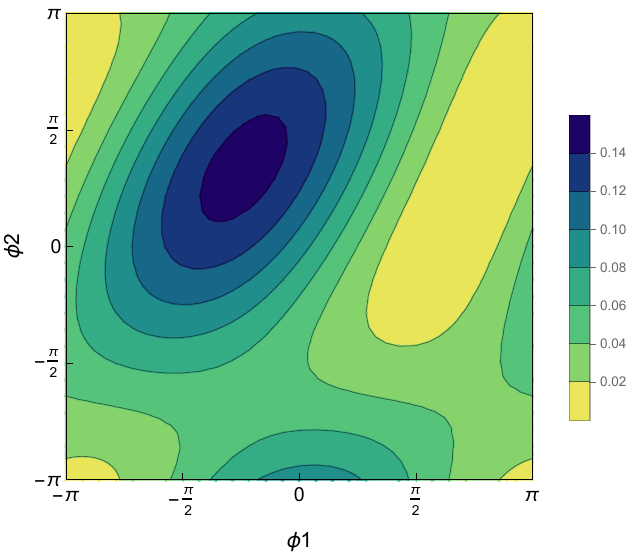}
		\includegraphics[width=0.45\linewidth]{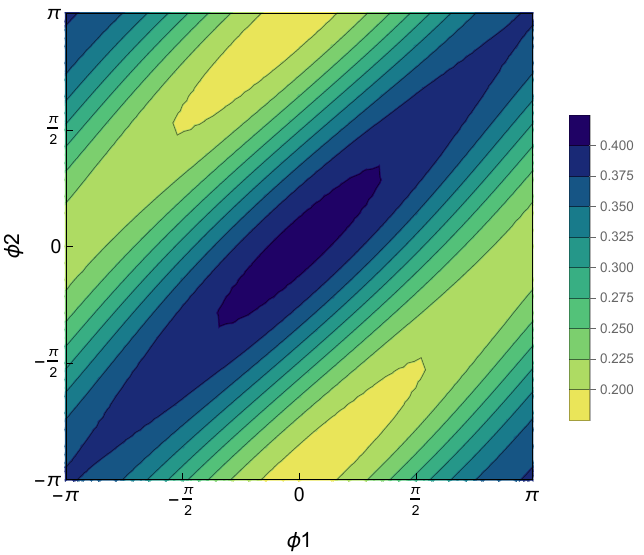}
		\includegraphics[width=0.45\linewidth]{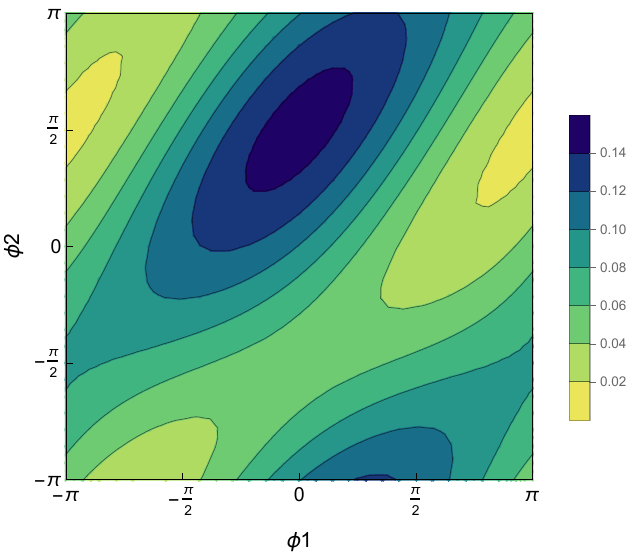}
	\end{center}
	\caption{Impact of Majorana phases $\phi_1,\phi_2$ on the branching ratio of $H^{++}\to \mu^+\mu^+$. The left two panels are for NH, while the right two panels are for IH. In the upper two panels, we fix $\delta_{\text{CP}}=\pi/2$. In the lower two panels, we fix $\delta_{\text{CP}}=\pi$. The lightest neutrino mass is fixed at 0.01 eV for both NH and IH. The other parameters are fixed to the best fit values \cite{Esteban:2020cvm}. }
	\label{BRmm}
\end{figure}

Because the Majorana phases $\phi_1,\phi_2$ can not be directly measured by the neutrino oscillation experiments, the branching ratio of $H^{++}\to \ell^+\ell^+$ provides an important pathway. In Figure \ref{BRmm}, we take BR($H^{++}\to \mu^+\mu^+$) for illustration. Typically, the theoretically predicted BR($H^{++}\to \mu^+\mu^+$) in NH is always larger than that in IH, which is helpful to  distinguish these two scenarios. The Dirac phase $\delta_{\text{CP}}$ has a relatively small impact on BR($H^{++}\to \mu^+\mu^+$). For certain measured values of BR($H^{++}\to \mu^+\mu^+$), the favored region is then determined in the $\phi_1-\phi_2$ plane. The individual values of $\phi_1,\phi_2$ might be obtained by the combination of other modes \cite{Akeroyd:2007zv}.

\section{Single Production of $H^{\pm\pm}$ at Muon Collider}\label{SC:PD}

\begin{figure}
	\begin{center}
		\includegraphics[width=0.45\linewidth]{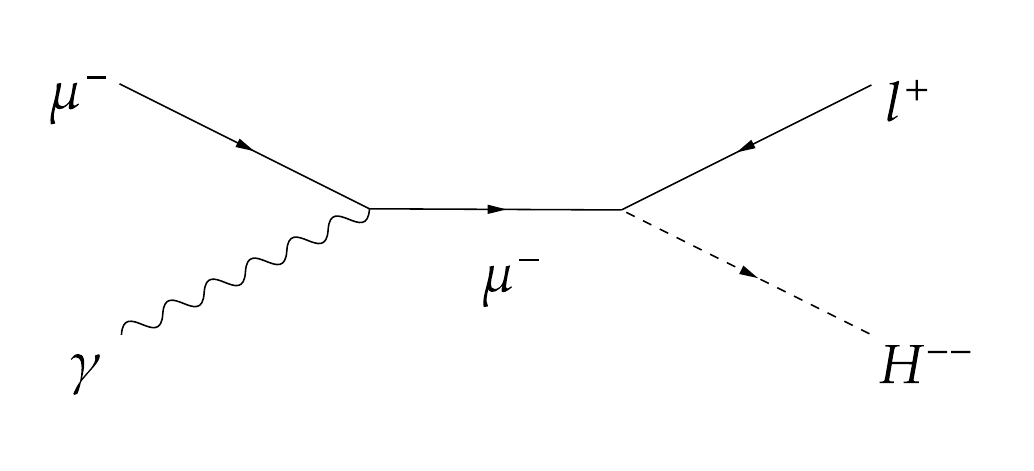}
		\includegraphics[width=0.45\linewidth]{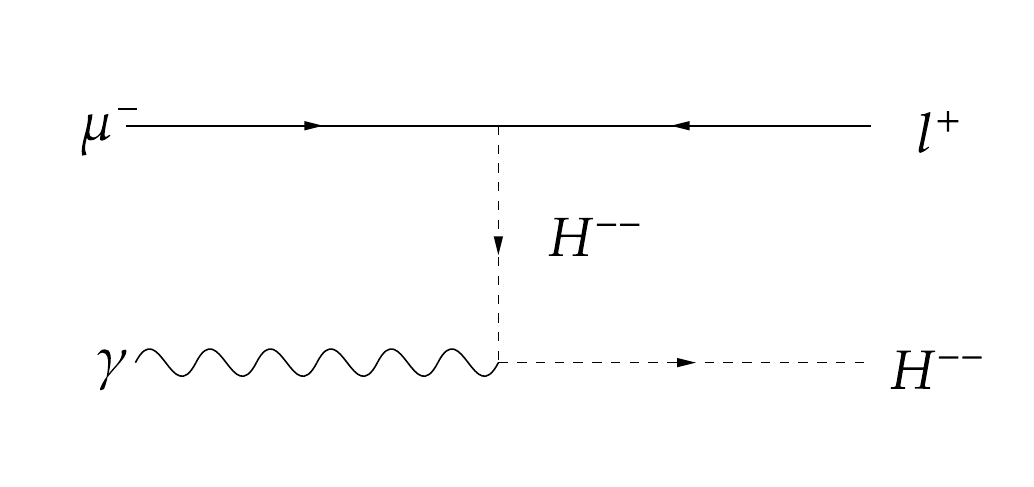}
		\includegraphics[width=0.45\linewidth]{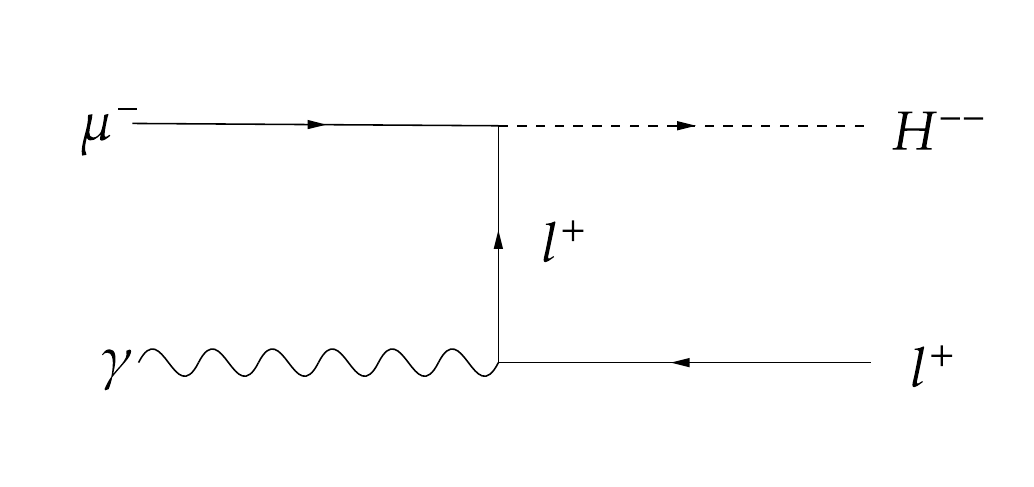}
	\end{center}
	\caption{Feynman diagram of $ \mu^- \gamma \to \ell^{+} H^{--}$. }
	\label{Feyn}
\end{figure}

\begin{figure}
	\begin{center}
		\includegraphics[width=0.8\linewidth]{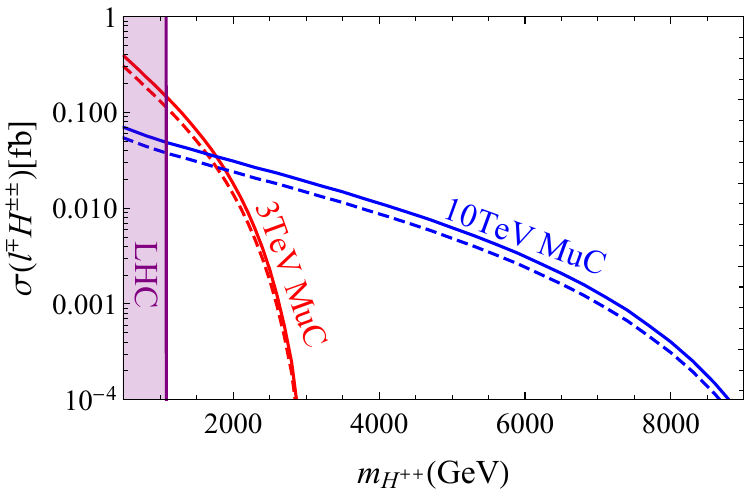}
	\end{center}
	\caption{Total cross section of $ \mu^\pm \gamma \to \ell^{\mp} H^{\pm\pm}(\ell = e,\mu,\tau)$ at muon collider. The purple region is excluded by the current LHC search \cite{ATLAS:2022pbd}. Solid lines are the results of NH, while dashed lines are the results of IH. In each scenario, the neutrino oscillation parameters are fixed to the best fit values in Ref \cite{Esteban:2020cvm}. The lightest neutrino mass is assumed to be 0.01 eV. The two Majorana phases are fixed to be zero. Here, we also consider $v_\Delta=1$ eV for illustration. }
	\label{CSHpp}
\end{figure}

Pair production of doubly charged Higgs is dominant by direct $\mu^+\mu^-$ annihilation or vector boson fusion processes at muon collider \cite{Li:2023ksw}, which depends on the gauge coupling of $H^{\pm\pm}$. In this paper, we consider the single production of doubly charged Higgs at muon collider. This process is produced via the subprocess of $\mu \gamma$ collision, which is shown explicitly in Figure \ref{Feyn}. The full process is
\begin{equation}
	\mu^+\mu^-\to \mu^+ \gamma \mu^- \to \mu^\mp \ell^\mp H^{\pm\pm},
\end{equation}
where the photon $\gamma$ emitted from the muon interacts with another muon. Then the cross section of single production involves the Yukawa coupling $h_{\mu \ell}$. The total cross section can be calculated as \cite{Yue:2010zu}
\begin{equation}
	\sigma(\mu^+\mu^-\to \mu^\mp \ell^\mp H^{\pm\pm}) = \int f_{\gamma/\mu}(x)\hat{\sigma}(\mu^\pm \gamma \to \ell^\mp H^{\pm \pm}) dx,
\end{equation}
where $f_{\gamma/\mu}(x)$ is the photon distribution function \cite{Budnev:1975poe,Han:2020uid,Ruiz:2021tdt}. In this paper, we use Madgraph5\_aMC@NLO \cite{Alwall:2014hca} to obtain the numerical results. The total cross section of $\mu^+\mu^-\to \mu^\mp \ell^\mp H^{\pm\pm}$ is shown in Figure \ref{CSHpp}. From Eqn. \eqref{Eq:hij}, it is clear that the Yukawa coupling depends on the neutrino oscillation parameters and triplet VEV $v_\Delta$. We have fixed the relevant parameters to illustrate the dependence of total cross section on $m_{H^{++}}$ in Figure \ref{CSHpp}.

The NH scenario has a slightly larger cross section than the IH scenario. Constrained by the threshold, the pair production channel could only probe $m_{H^{++}}<\sqrt{s}/2$ \cite{Li:2023ksw}, while the single production channel  in principle could test $m_{H^{++}}<\sqrt{s}$. At the 3 TeV muon collider,  $\sigma(\mu^+\mu^-\to \mu^\mp \ell^\mp H^{\pm\pm})$ is larger than $10^{-3}$~fb when $m_{H^{++}}\lesssim2500$ GeV. With an integrated luminosity of $1000~\text{fb}^{-1}$, the single production channel may further discover the range of  $m_{H^{++}}\in[1500,2000]$ GeV at the 3 TeV muon collider with more than ten events. When $m_{H^{++}}<1700$ GeV, the total cross section at the 3 TeV muon collider is larger than that at the 10 TeV energy. For the 10 TeV muon collider with $10~\text{ab}^{-1}$ data, there still could be a few events when $m_{H^{++}}\sim8000$ GeV. It should be noted that the cross section heavily depends on the triplet VEV $v_\Delta$. As $h_{\mu\ell}\propto (v_\Delta)^{-1}$, decreasing $v_{\Delta}$ will increase the cross section, which then enlarges the experimental reach. The sensitive region in the $v_\Delta-m_{H^{++}}$ plane will be investigated in Section \ref{SC:SG} after taking into account the SM background and the acceptance cut efficiency.

\begin{figure}
	\begin{center}
		\includegraphics[width=0.45\linewidth]{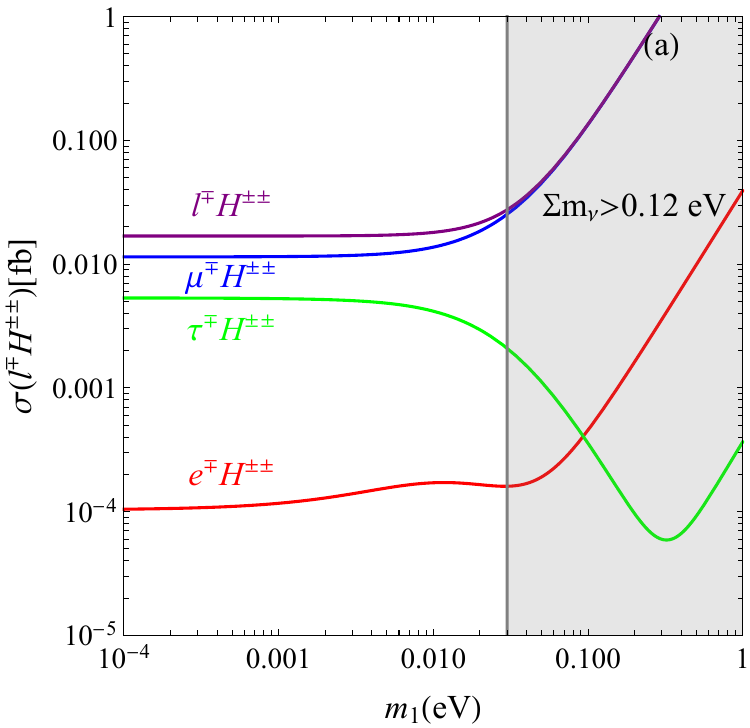}
		\includegraphics[width=0.45\linewidth]{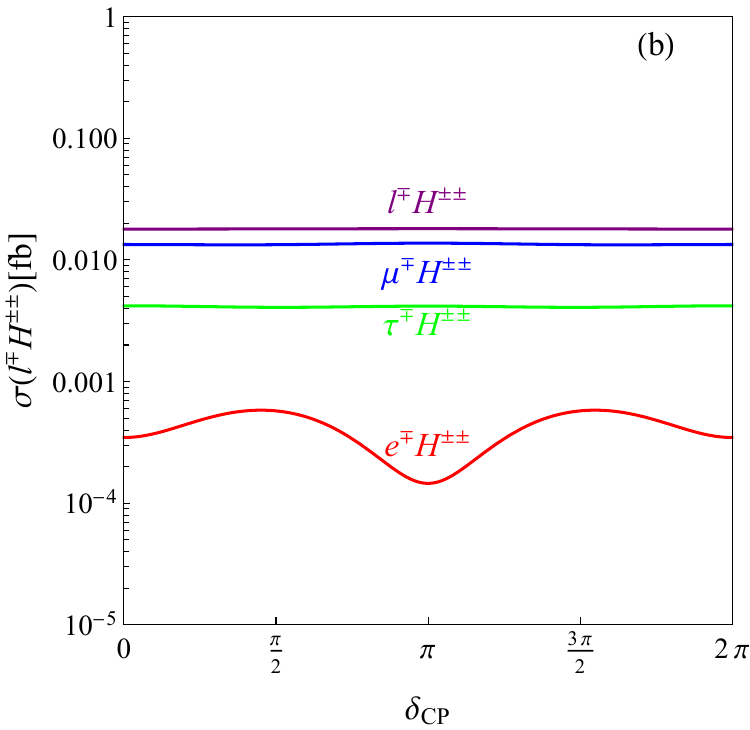}
		\includegraphics[width=0.45\linewidth]{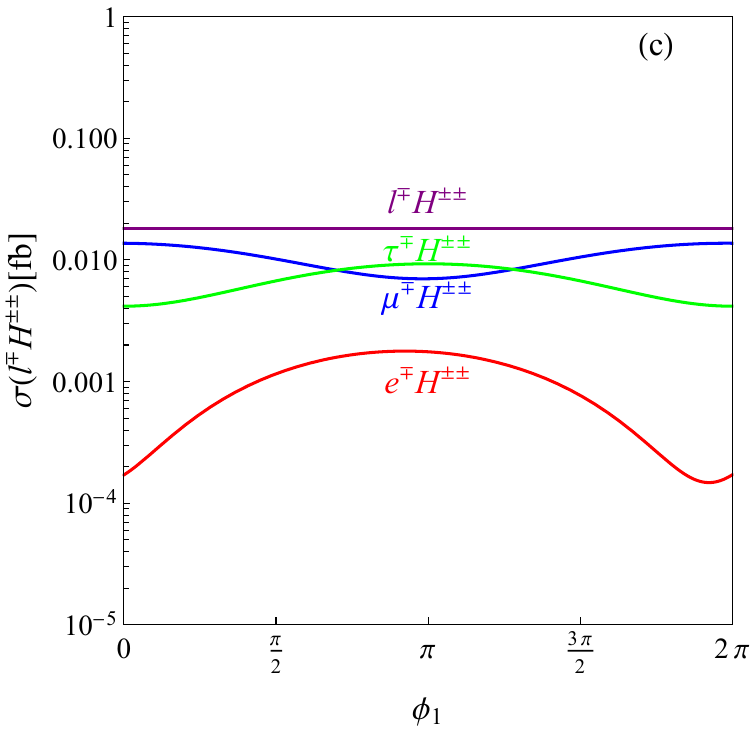}
		\includegraphics[width=0.45\linewidth]{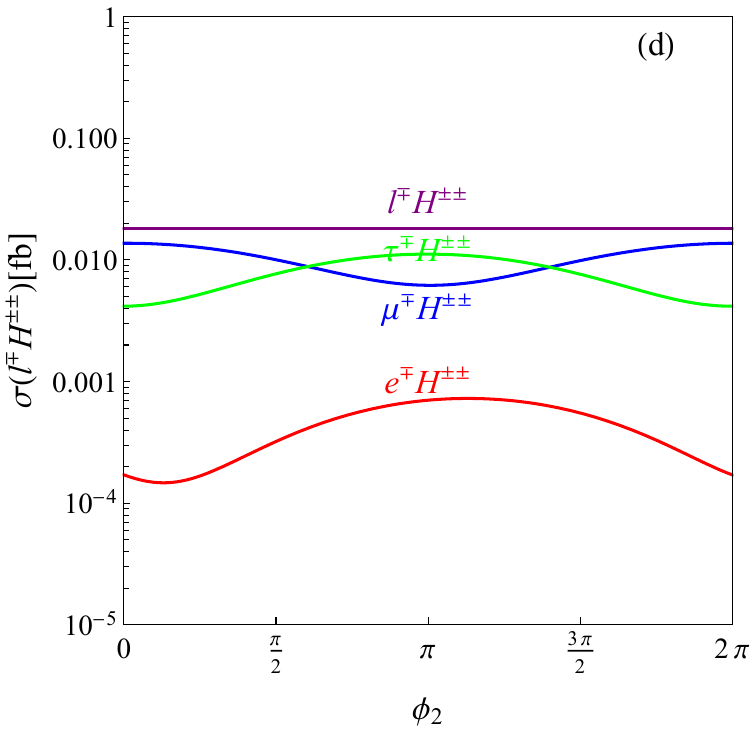}
	\end{center}
	\caption{Cross section of $ \ell^{\mp} H^{\pm\pm}$ at th 3 TeV muon collider for $m_{H^{++}}=2000$ GeV in NH scenario. The neutrino oscillation parameters are fixed to the best fit values in Ref \cite{Esteban:2020cvm}.  Here, we also assume $v_\Delta=1$ eV for illustration. The purple lines correspond  to the total cross sections with the sum of all three channels. In panel (a), we fix $\delta_{\text{CP}}=197^{\circ}, \phi_1=\phi_2=0$. The gray region is excluded by the Planck limit $\sum m_\nu <0.12$ eV \cite{Planck:2018vyg}. In panel (b), we fix $m_1=0.01$ eV, $\phi_1=\phi_2=0$. In panel (c), we fix $m_1=0.01$ eV, $\delta_{\text{CP}}=197^{\circ}$, $\phi_2=0$. In panel (d), we fix $m_1=0.01$ eV, $\delta_{\text{CP}}=197^{\circ}$, $\phi_1=0$. }
	\label{CSNH}
\end{figure}

We then consider the dependence of the cross section on the neutrino oscillation parameters. Here, we take $m_{H^{++}}=2000$ GeV at the 3 TeV muon collider for illustration. Dependence of neutrino oscillation parameters are quite similar for other values of $m_{H^{++}}$ and collision energy, thus they will not be repeated here. The results for the NH scenario are shown in Figure \ref{CSNH}. The dominant production channel is $\mu^\mp H^{\pm\pm}$ for $\phi_1=\phi_2=0$. Meanwhile, the $e^\mp H^{\pm\pm}$ channel is about $10^{-4}$ fb, which is hard to detect at the muon collider. For the lightest neutrino mass $m_1$, we find that the total cross section $\sigma(\ell^\mp H^{\pm\pm})$ increases with larger $m_1$. However, under the cosmological constraint $\sum m_\nu <0.12$~eV, such dependence is not obvious when $m_1\lesssim0.01$ eV. The Dirac phase $\delta_{\text{CP}}$ has a significant impact on the $e^\mp H^{\pm\pm}$ channel, but the overall cross section is always too small. The dependence of $\mu^\mp H^{\pm\pm}$ and $\tau^\mp H^{\pm\pm}$ channel on $\delta_{\text{CP}}$ is very weak, which leads to the total cross section  $\sigma(\ell^\mp H^{\pm\pm})$ not sensitive to $\delta_{\text{CP}}$. Both the Majorana phase $\phi_1$ and $\phi_2$ could greatly alert the individual production channel. By measuring the cross section ratio  $\sigma(\mu^\mp H^{\pm\pm})/\sigma(\tau^\mp H^{\pm\pm})$, the Majorana phases can also be determined.  Notice that the  total cross section $\sigma(\mu^+\mu^-\to \mu^\mp \ell^\mp H^{\pm\pm})\propto\sum |h_{\mu\ell}|^2\propto(m_\nu^*m_\nu)_{\mu \mu}$, which is independent of Majorana phases $\phi_1$ and $\phi_2$. This result is also true for the IH scenario.

\begin{figure}
	\begin{center}
		\includegraphics[width=0.45\linewidth]{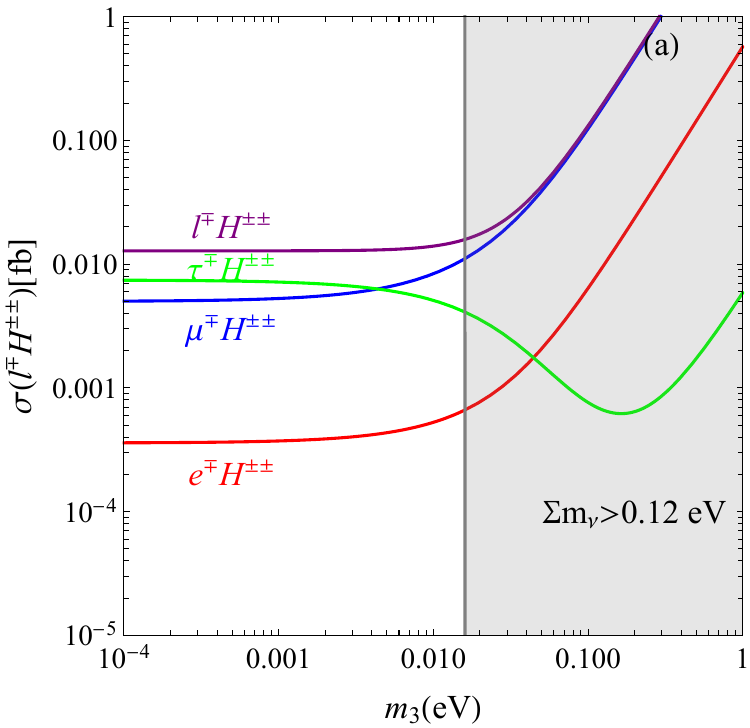}
		\includegraphics[width=0.45\linewidth]{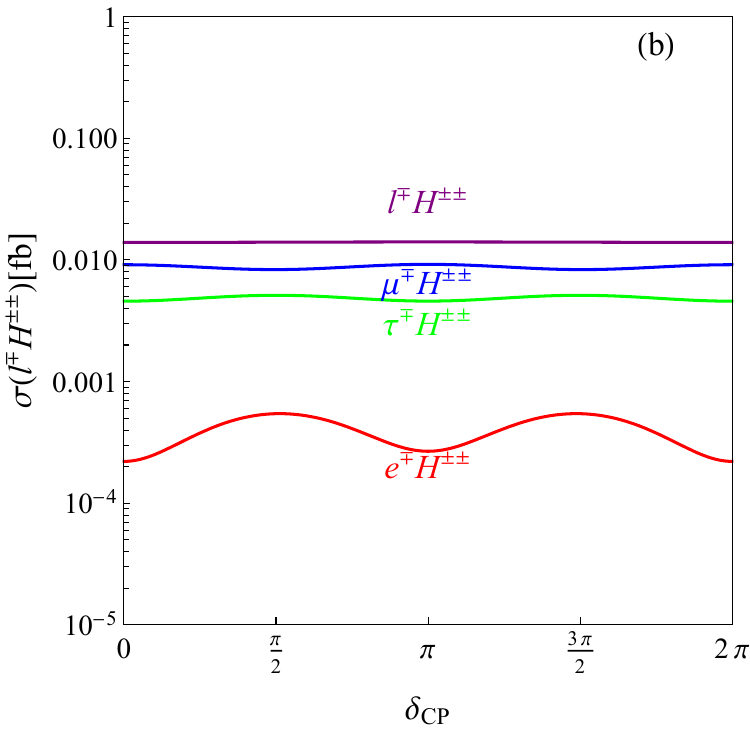}
		\includegraphics[width=0.45\linewidth]{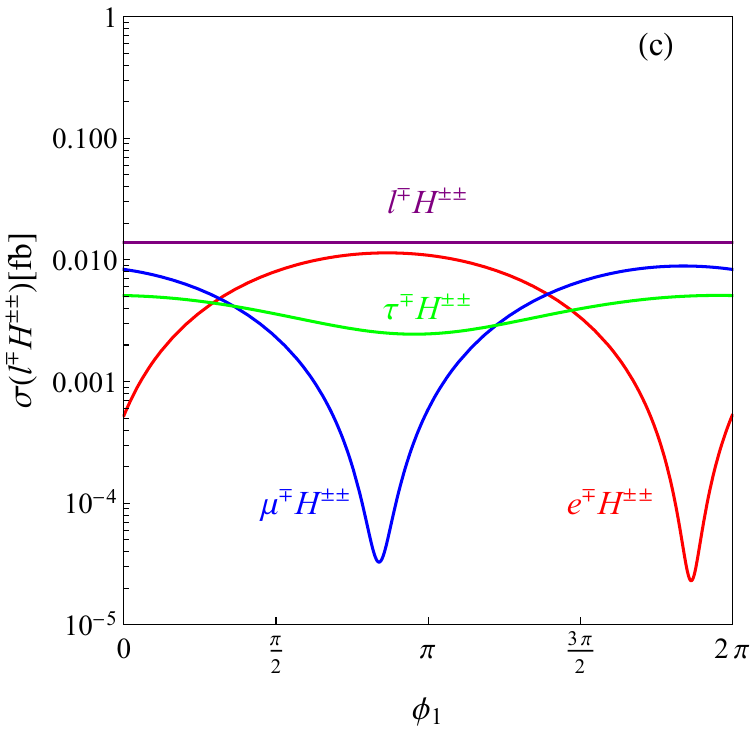}
		\includegraphics[width=0.45\linewidth]{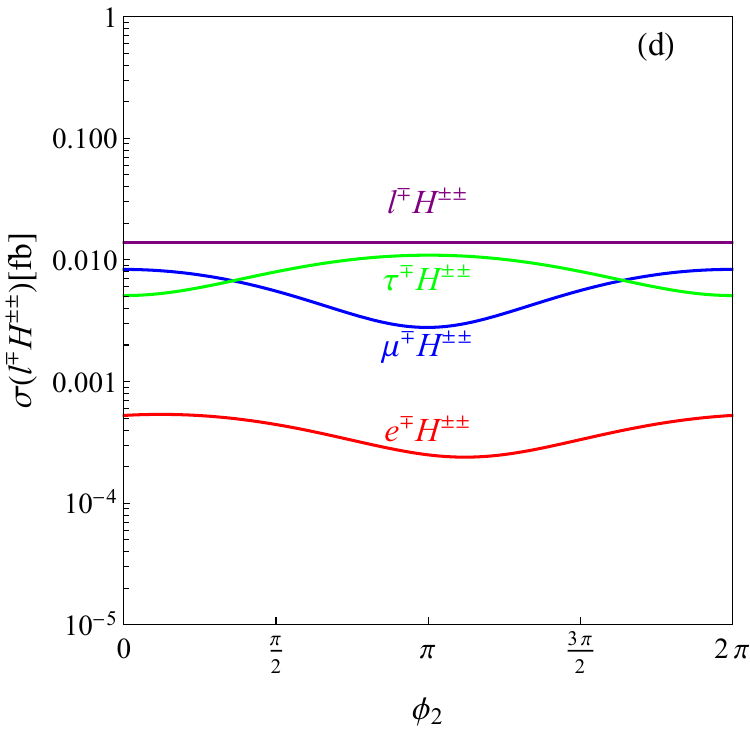}
	\end{center}
	\caption{Same as Figure \ref{CSNH}, but for the IH scenario. In panel (a), we fix $\delta_{\text{CP}}=282^{\circ}, \phi_1=\phi_2=0$. The gray region is excluded by the Planck limit $\sum m_\nu <0.12$ eV \cite{Planck:2018vyg}. In panel (b), we fix $m_1=0.01$ eV, $\phi_1=\phi_2=0$. In panel (c), we fix $m_1=0.01$ eV, $\delta_{\text{CP}}=282^{\circ}$, $\phi_2=0$. In panel (d), we fix $m_1=0.01$ eV, $\delta_{\text{CP}}=282^{\circ}$, $\phi_1=0$. }
	\label{CSIH}
\end{figure}

The results for the IH scenario are shown in Figure \ref{CSIH}. With $\delta_{\text{CP}}=282^{\circ},\phi_1=\phi_2=0$, the cross section of $\tau^\mp H^{\pm\pm}$ channel becomes larger than the $\mu^\mp H^{\pm\pm}$ channel when $m_3<0.004$ eV. The total cross section $\sigma(\ell^\mp H^{\pm\pm})$ depends slightly on the Dirac phase $\delta_{\text{CP}}$. For $m_3=0.01$ eV and $\delta_{\text{CP}}=282^{\circ}$, the Majorana phase $\phi_1$ has a larger impact on the individual cross section than the Majorana phase $\phi_2$. Especially for $\phi_1\sim0.8\pi$, the $e^\mp H^{\pm\pm}$ channel becomes the dominant channel, while the $\mu^\mp H^{\pm\pm}$ channel is heavily suppressed.  

\section{Signature at Muon Collider}\label{SC:SG}

\begin{figure}
	\begin{center}
		\includegraphics[width=0.45\linewidth]{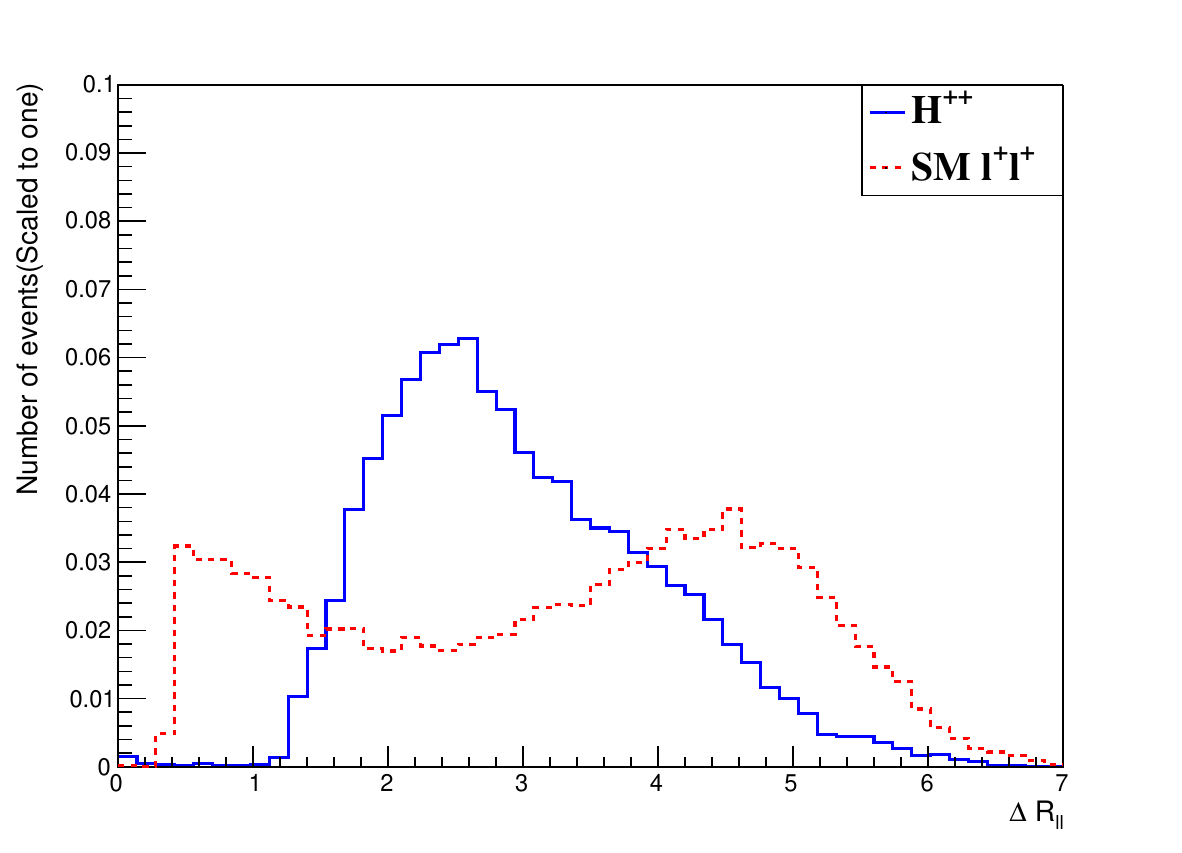}
		\includegraphics[width=0.45\linewidth]{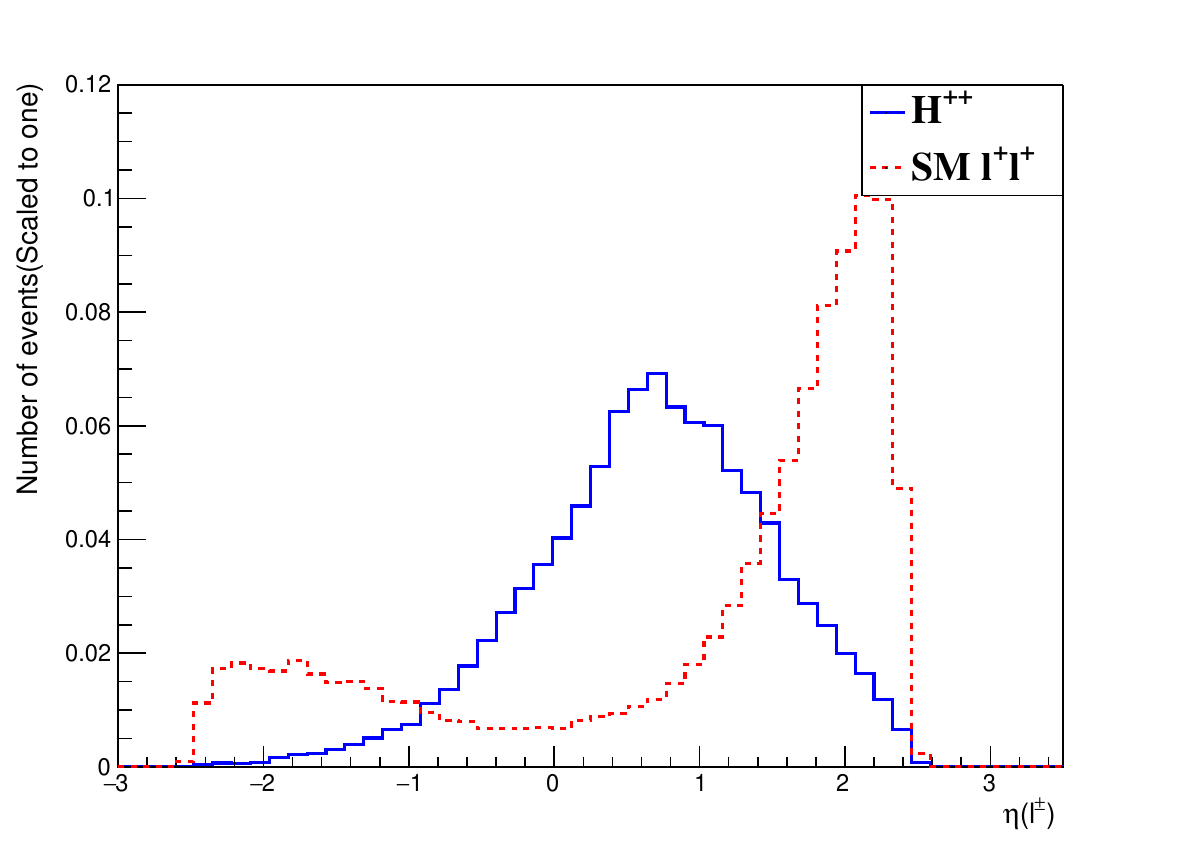}
		\includegraphics[width=0.45\linewidth]{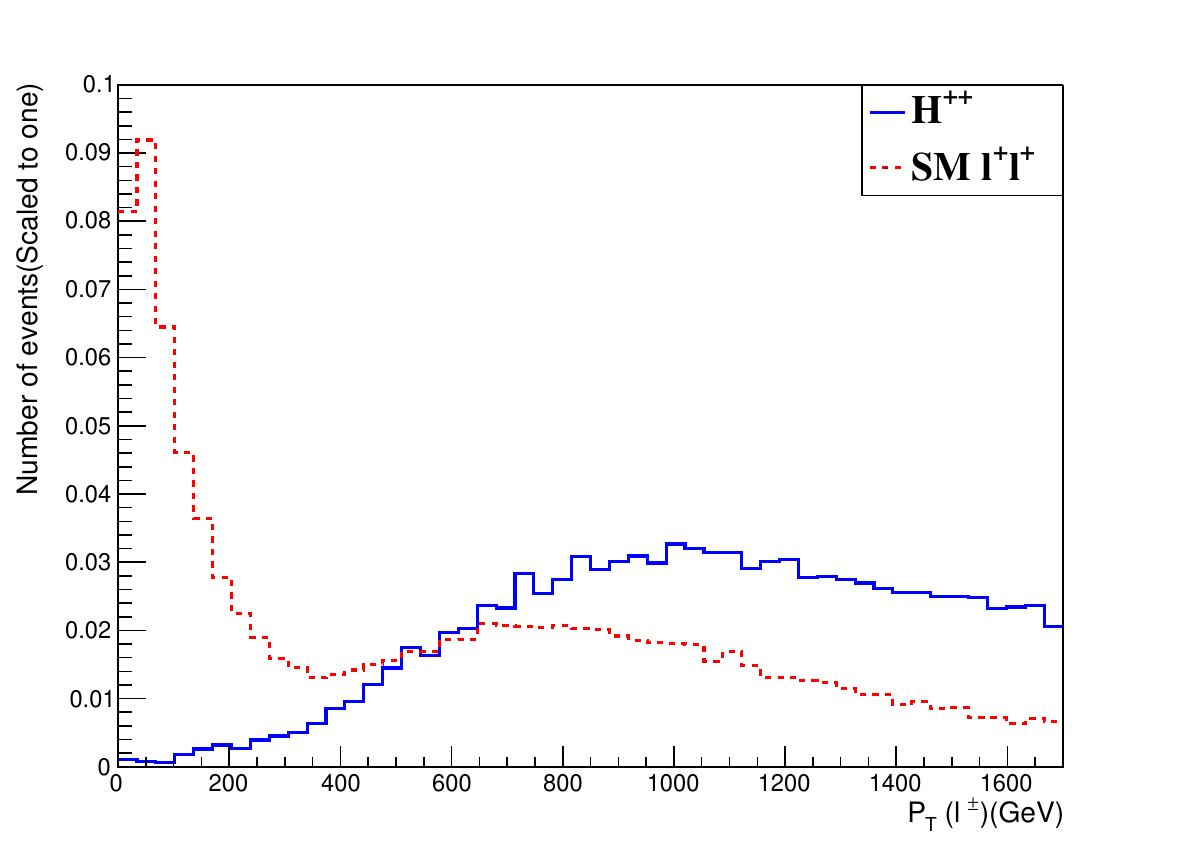}
		\includegraphics[width=0.45\linewidth]{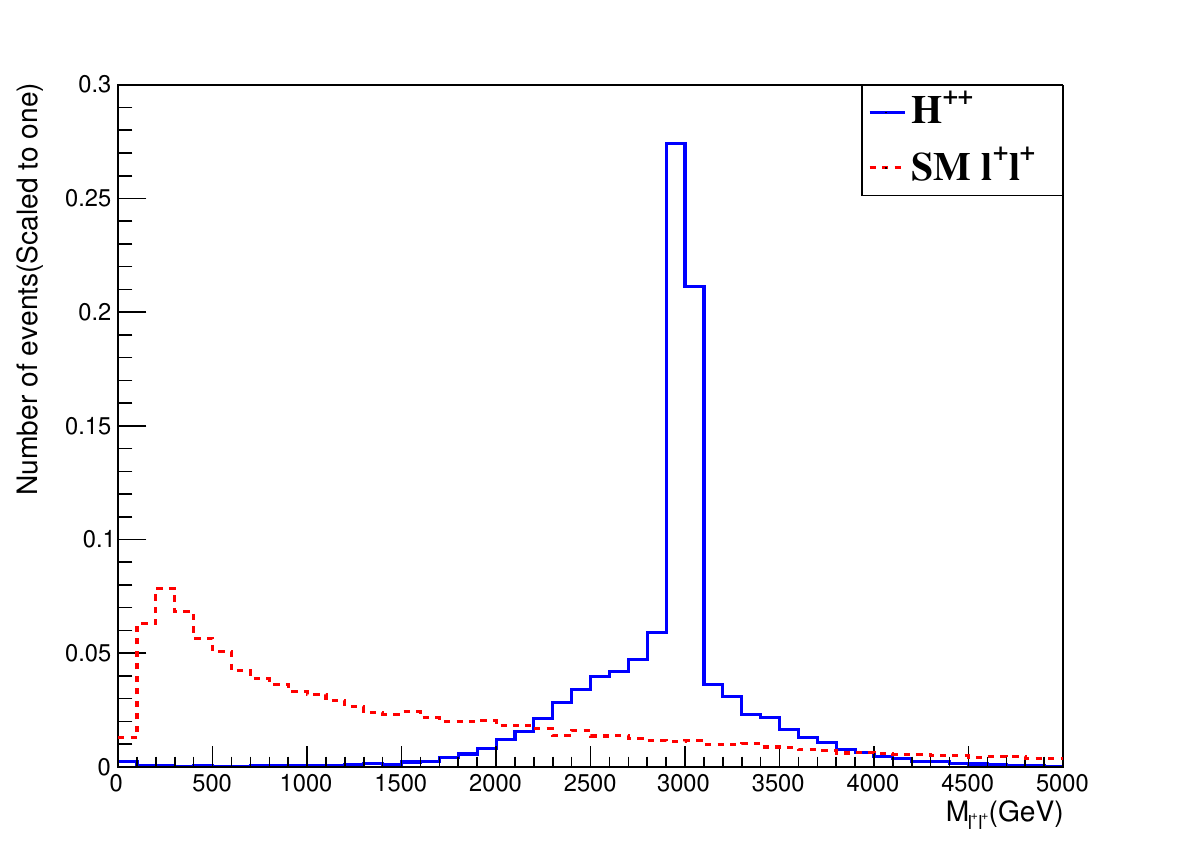}
	\end{center}
	\caption{ Distribution of separation $\Delta R_{\ell \ell}=\sqrt{\Delta \eta_{\ell^\pm\ell^\pm}^2+\Delta \phi_{\ell^\pm\ell^\pm}^2}$, pseudorapidity $\eta(\ell^\pm)$, transverse momenta $P_T(\ell^\pm)$, and invariant mass $m_{\ell^\pm\ell^\pm}$ at the 10 TeV muon collider. }
	\label{MCDT}
\end{figure}

After the single production of doubly charged scalar at muon collider, the same sign dilepton decay channel leads to the signature
\begin{equation}
	\mu^+\mu^-\to \mu^\mp \ell^{\prime\mp} + H^{\pm\pm} (\to \ell^\pm\ell^\pm),
\end{equation}
where $\ell^{\prime}=e,\mu,\tau$, and $\ell=e,\mu$ in the following analysis. In this paper, we consider the inclusive same sign dilepton signature $\ell^\pm\ell^\pm+X$, where $X$ denotes the undetected charged leptons. Due to lower tagging efficiency, the $\tau$ final states are not taken into account. The dominant SM background comes from 
\begin{equation}
	\mu^+\mu^-\to \ell^{+}\ell^{-} \ell^+ \ell^-.
\end{equation}
The parton-level events for both signal and background are generated by Madgraph5\_aMC@NLO \cite{Alwall:2014hca}. Pythia8 \cite{Sjostrand:2007gs} is used for parton showering and hadronization. Finally, the detector simulation is performed by Delphes3 \cite{deFavereau:2013fsa} with the detector card of the muon collider.

\begin{figure}
	\begin{center}
		\includegraphics[width=0.45\linewidth]{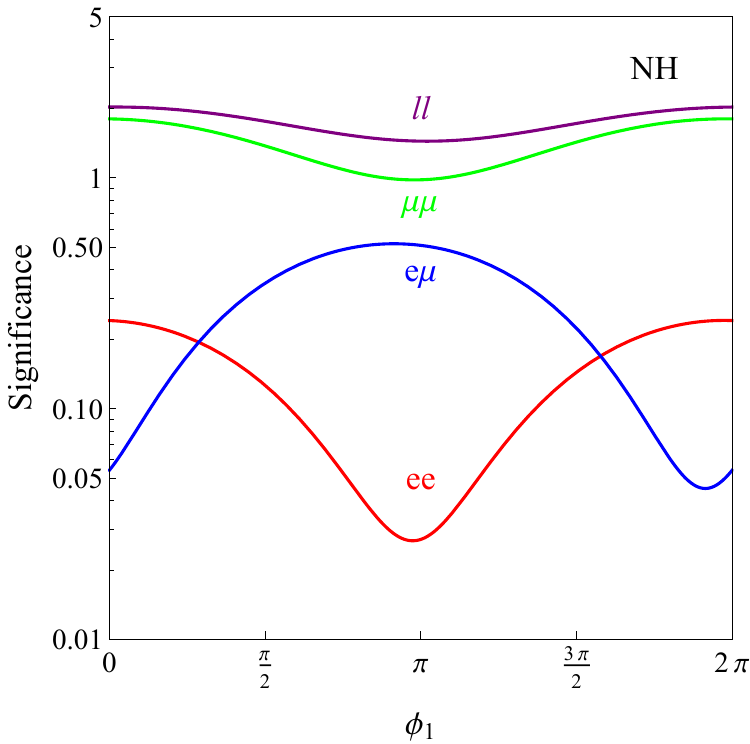}
		\includegraphics[width=0.45\linewidth]{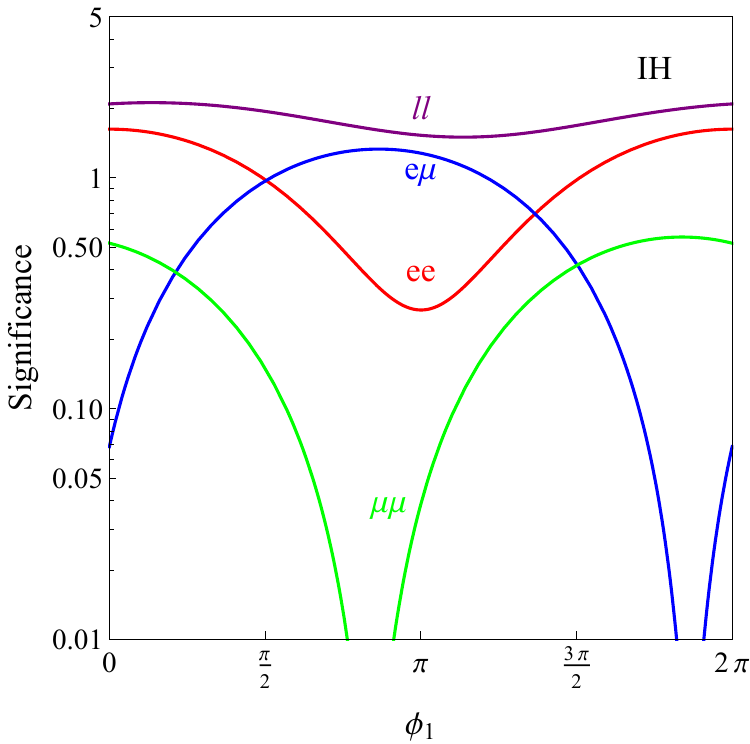}
		\includegraphics[width=0.45\linewidth]{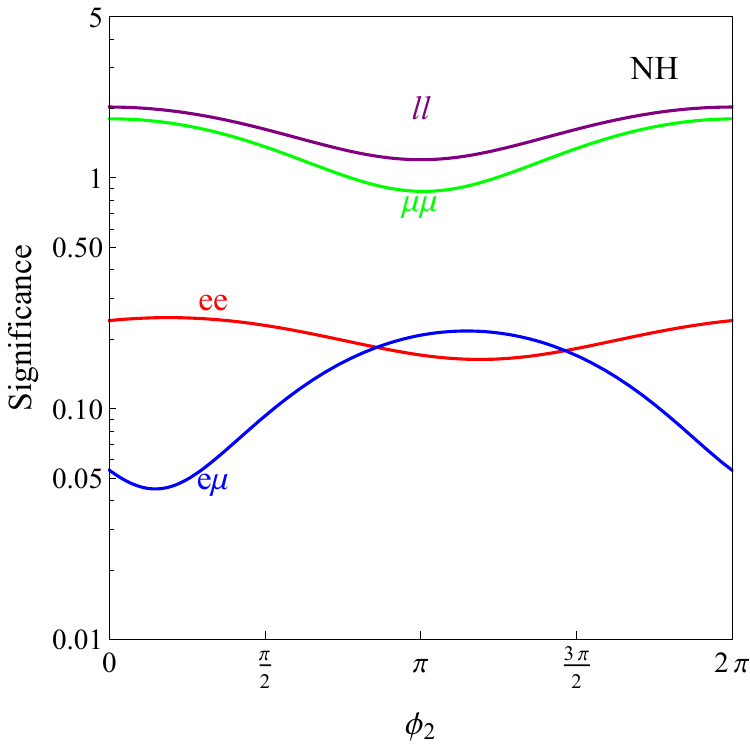}
		\includegraphics[width=0.45\linewidth]{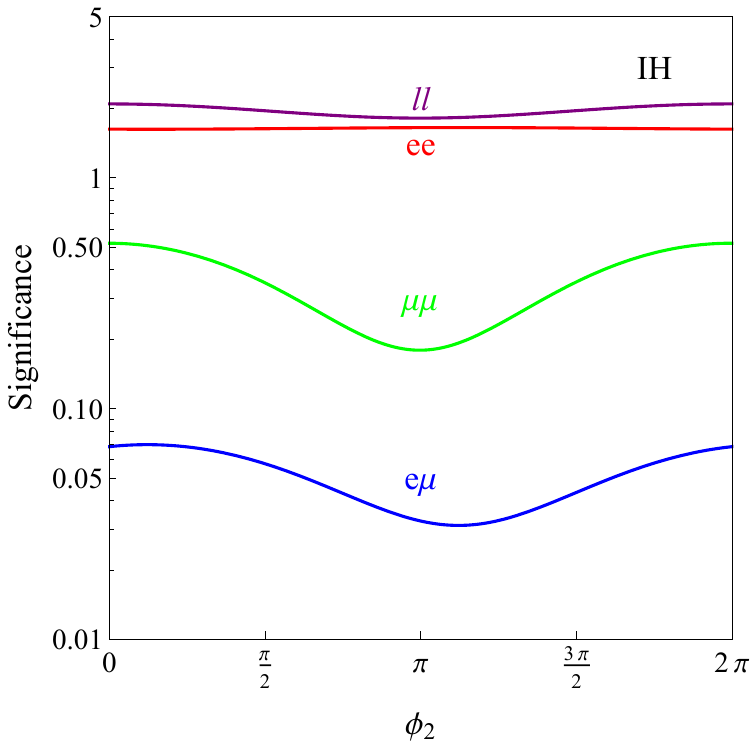}
	\end{center}
	\caption{Significance of the same sign dilepton signature at the 3 TeV muon collider. Here, we fix $m_{H^{\pm\pm}}=2000$~GeV and $v_\Delta=1$ eV. The lightest neutrino mass is fixed as $m_{1,3}=0.01$ eV, while the other oscillation parameters are fixed to the best fit values \cite{Esteban:2020cvm}. }
	\label{SGPhi}
\end{figure}

After requiring  the events with two same sign leptons $N(\ell^\pm)=2$, we apply the following basic cuts for the leptons in the final states
\begin{equation}
	\Delta R_{\ell\ell}>0.4, |\eta(\ell^\pm)|<2.5, P_T(\ell^\pm) > 50~\text{GeV}.
\end{equation}
In Figure \ref{MCDT}, some variables for the same sign dilepton signature $\ell^\pm\ell^\pm$ after basic cuts are shown. Based on these distributions, we then tighten the selection cuts as
\begin{equation}
	1.2<\Delta R_{\ell\ell}<5.4, |\eta(\ell^\pm)|<2.0, P_T(\ell^\pm) > 350~\text{GeV}.
\end{equation}
As the same sign dilepton come from the decay of $H^{\pm\pm}$, there is a resonance peak around $m_{H^{\pm\pm}}$. The background can be further suppressed by the invariant mass cut
\begin{equation}
	|m_{\ell^\pm\ell^\pm}-m_{H^{\pm\pm}}|<m_{H^{\pm\pm}}/5.
\end{equation}

The significance is calculated as \cite{Cowan:2010js}
\begin{equation}
	\mathcal{S}=\sqrt{2\left[(N_S+N_B)\ln\left(1+\frac{N_S}{N_B}\right)-N_S\right]},
\end{equation}
where $N_S$ and $N_B$ are the event number of signal and background after cuts respectively. In this paper, we assume the integrated luminosity of the 3 TeV and 10 TeV muon collier to be $1~\text{ab}^{-1}$ and $10~\text{ab}^{-1}$ respectively \cite{Delahaye:2019omf}.

According to the results in Section \ref{SC:DC}, the Majorana phases have a great impact on the branching ratio of $H^{\pm\pm}$. In Figure \ref{SGPhi}, we show the significance as a function of Majorana phases. For both NH and IH scenarios, the significance is about 2  in combination with the $\mu\mu,e\mu,ee$ channels for $m_{H^{\pm}}=2000$ GeV and $v_\Delta=1$ eV at the 3 TeV muon collider. For the NH scenario, the $\mu\mu$ channel is always the most promising channel, while  the lepton flavor violation $e\mu$ channel is important to probe the Majorana phases. For the IH scenario, the dominant channel heavily depends on the Majorana phase $\phi_1$. When $\phi_1=0$, the $ee$ channel is the dominant one, while the $e\mu$ channel becomes the dominant for $\phi_1\sim\pi$.

\begin{figure}
	\begin{center}
		\includegraphics[width=0.45\linewidth]{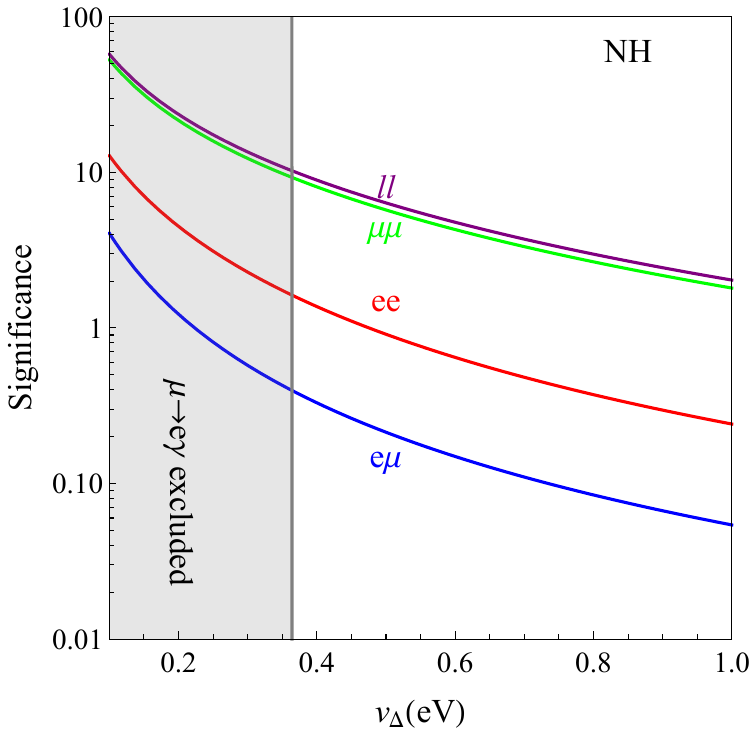}
		\includegraphics[width=0.45\linewidth]{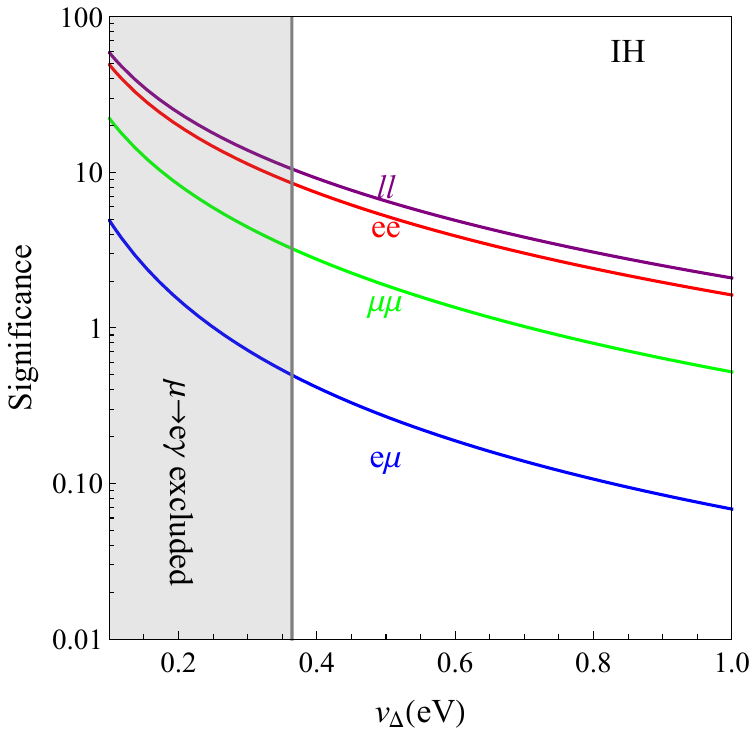}
	\end{center}
	\caption{Significance of the same sign dilepton signature at the 3 TeV muon collider. Here, we assume $m_{H^{\pm\pm}}=2000$~GeV, $m_{1,3}=0.01$ eV, $\phi_1=\phi_2=0$,  while the other oscillation parameters are fixed to the best fit values \cite{Esteban:2020cvm}. The gray region is excluded by $\mu\to e\gamma$.}
	\label{SGVd}
\end{figure}

Another important parameter is the triplet VEV $v_\Delta$. In Figure \ref{SGVd}, we show the significance  as a function of $v_\Delta$. As $h_{ij}\propto v_{\Delta}^{-1}$, decreasing $v_\Delta$ will increase the Yukawa coupling $h_{ij}$, thus the signal significance. For $m_{H^{\pm\pm}}=2000$ GeV, the significance would exceed $5\sigma$ when $v_\Delta\lesssim0.6$ eV. With vanishing Majorana phases, the significance of the $ee$ channel in NH is less than $2\sigma$ under the current limit. The significance of the $\mu\mu$ channel could reach $4\sigma$ in IH. The flavor violation $e\mu$ channel is unpromising in Figure \ref{SGVd}, but could becomes promising by modifying the Majorana phases $\phi_{1,2}$.

\begin{figure}
	\begin{center}
		\includegraphics[width=0.8\linewidth]{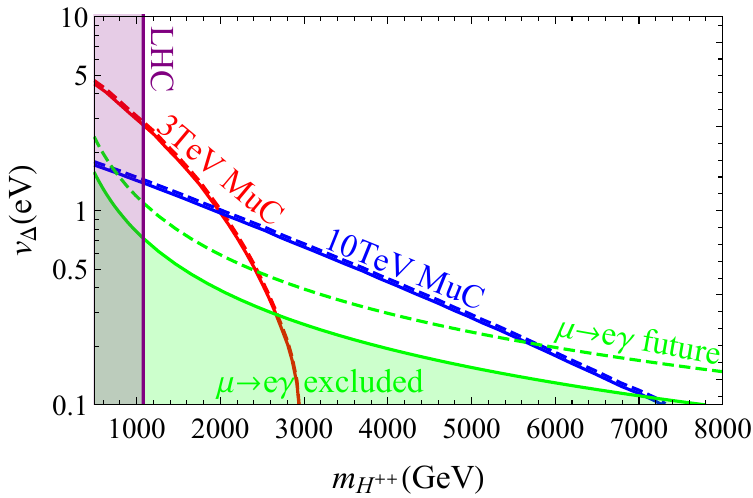}
	\end{center}
	\caption{The $2\sigma$ sensitivity of the same sign dilepton signature from single production of doubly charged scalar at muon collider. Solid lines are the results of NH, while dashed lines are the results of IH. The purple region is excluded by the current LHC search \cite{ATLAS:2022pbd}. The green region is excluded by the current $\mu\to e\gamma$ \cite{MEG:2016leq}. The dashed green line is the result of the future BR$(\mu\to e\gamma)<6\times10^{-14}$ limit \cite{Baldini:2013ke}. Here, we assume the lightest neutrino mass $m_{1,3}=0.01$ eV and the Majorana phases $\phi_1=\phi_2=0$.}
	\label{Sens}
\end{figure}

We then investigate the sensitive region of the same sign dilepton signature in the $v_\Delta-m_{H^{\pm\pm}}$ parameter space, which is shown in Figure \ref{Sens}. Both NH and IH scenarios lead to quite similar results. This is because that the NH scenario has a larger production cross section of $\mu^+\mu^-\to \mu^\mp \ell^\mp H^{\pm\pm}$, while the IH scenario has a larger dilepton branching ratio of $H^{\pm\pm}\to \ell^\pm\ell^{\pm}(\ell=e,\mu)$. Under the current $\mu\to e\gamma$ limit, the 3 TeV muon collider could probe $m_{H^{++}}\lesssim2.6$ TeV, while the 10 TeV muon collider would extend the mass to about $7.1$ TeV. Certain region with $v_\Delta\sim0.5$ eV is also in the reach of future $\mu\to e\gamma$ experiment, thus this region can be confirmed by both lepton flavor violation and collider. However, if no $\mu\to e\gamma$ is observed in the future, the sensitive region will be down to 5.7 TeV.

\section{Conclusion}\label{SC:CL}

The type-II seesaw is one of the most appealing pathways to explain the tiny neutrino masses. This model introduces a scalar triplet $\Delta$ with hypercharge $Y=2$, thus one distinct feature is the doubly charged Higgs $H^{\pm\pm}$. A small triplet VEV $v_\Delta$ is induced via the lepton number violation term $\mu \Phi^{\mathrm{T}} i\tau_{2} \Delta^{\dagger} \Phi$. Then tiny neutrino masses are generated through the Yukawa interaction  $h \overline{L_{L}^{c}} i \tau_{2} \Delta L_{L}$, which indicates that the Yukawa coupling $h$ is closely related to the neutrino oscillation parameters.

In this paper, we study the single production of doubly charged Higgs $H^{\pm\pm}$ in the type-II seesaw at the high energy muon collider. The single production channel is produced via the subprocess of $\mu \gamma$ collision as $\mu^+\mu^-\to \mu^+ \gamma \mu^- \to \mu^\mp \ell^\mp H^{\pm\pm}$. This process depends on the Yukawa coupling $h$, hence the neutrino oscillation parameters. We then study the impact of the lightest neutrino mass $m_{1,3}$, Dirac phase $\delta_{\text{CP}}$, and Majorana phases $\phi_{1,2}$. Under the cosmological limit, the lightest neutrino mass $m_{1,3}$ has a relatively small impact on the cross section. The Dirac phase $\delta_{\text{CP}}$ has a great impact on the $e^{\mp}H^{\pm\pm}$ channel, but the corresponding cross section might be too small to be detected. Meanwhile, the Majorana phases $\phi_{1,2}$ could significantly alert the individual cross section, although the total cross section is not.

One advantage of the single production channel is the ability to probe doubly charged Higgs above the mass threshold $m_{H^{\pm\pm}}>\sqrt{s}/2$.  After simulating the signal and background, we find that the same sign dilepton signature from $H^{\pm\pm}\to \ell^\pm\ell^\pm$ could probe $m_{H^{\pm\pm}}\lesssim2.6(7.1)$ TeV at the 3 (10) TeV muon collider when the triplet VEV $v_\Delta\lesssim3$ eV. The Majorana phases $\phi_{1,2}$ also have great impact on the significance of the same sign dilepton signature by modifying the branching ratio of $H^{\pm\pm}\to\ell^\pm\ell^\pm$. For the NH scenario, the $\mu\mu$ channel is the most promising channel. For the IH scenario, the $ee$ channel is the most promising one when $\phi_1=\phi_2=0$, while the $e\mu$ channel becomes the dominant when $\phi_1\sim\pi,\phi_2=0$. Therefore, the Majorana phases $\phi_{1,2}$ can be determined by precise measurement of the branching ratio of $H^{\pm\pm}\to\ell^\pm\ell^\pm$.

\section*{Acknowledgments}
This work is supported by the National Natural Science Foundation of China under Grant No. 11805081, Natural Science Foundation of Shandong Province under Grant No. ZR2019QA021 and No. ZR2022MA056, the Open Project of Guangxi Key Laboratory of Nuclear Physics and Nuclear Technology under Grant No. NLK2021-07.


\end{document}